\documentstyle[aps,prl,multicol,epsfig]{revtex}
\begin{document}
\draft
\preprint{submitted to Phys. Rev. Lett.}
\title{
Quantum Information in Semiconductors: Noiseless Encoding \\ 
in a Quantum-Dot Array
}
\author{Paolo Zanardi$^{1,2}$ and Fausto Rossi$^{1,3}$}
\address{
$^{1}$ Istituto Nazionale per la Fisica della Materia (INFM) \\
$^{2}$ Institute for Scientific Interchange  Foundation, \\Villa Gualino,
Viale Settimio Severo 65, I-10133 Torino, Italy\\
$^{3}$ Dipartimento di Fisica, 
Universit\`a di Modena, 
Via G. Campi 213/A, 
I-41100 Modena, Italy \\ 
}
\date{\today}
\maketitle
\begin{abstract}

A potential implementation of quantum-computation schemes in 
semiconductor-based structures is proposed. In particular, an array of 
quantum dots is shown to be an  ideal quantum register for a 
noiseless information encoding. In addition to the suppression of 
phase-breaking processes in quantum dots due to the well-known phonon 
bottleneck, we show that a proper quantum encoding allows to realise 
a decoherence-free evolution on a time-scale long 
compared to the femtosecond scale of modern ultrafast laser 
technology. This result
 might
open the way to the realization of semiconductor-based quantum processors.

\end{abstract}
\begin{multicols}{2}
\narrowtext

The physical implementation of any computing device taking actual advantage
from the additional power provided by Quantum theory \cite{QC}
is extremely demanding.
In principle one should be able to perform, on a system with a well-defined 
state space,
long coherent quantum manipulations ({\sl gating}),
 precise quantum-state sinthesis and detection as well.
Ever since  the very beginning it has been recognized
that the major obstacle arises from the unavoidable open character
of any realistic quantum system.
The coupling with external (i.e. non computational) degrees of freedom
spoils the unitary structure of quantum evolution, which is the crucial
ingredient in quantum computation (QC).
This is the well-known decoherence problem \cite {DECO}.
The possibility to partly overcome such a difficulty by means of the {\sl
active stabilization } pursued by quantum error correction   is a definite
success of theoretical QC \cite{ERROR}.
Nevertheless, mostly due to the necessity of low decoherence  rates, 
the up-to-date proposals for experimental realizations of quantum processors
are based on quantum optics as well as atomic and molecular systems 
\cite{QC}.
Indeed, the extremely advanced technology in these fields
allows for the manipulations required in simple QC's.
It is however generally believed that  future applications, 
if any, of quantum 
information may hardly be realized in terms of such  systems, 
which do not permit the large-scale integration  of existing microelectronics 
technology.
In contrast, in spite of the serious difficulties related to the ``fast'' 
decoherence times, a solid-state implementation of QC
 seems to be the only way to benefit synergetically from the 
recent progress in ultrafast optoelectronics \cite{CCC} 
as well as in nanostructure fabrication and characterization \cite{QD-reviews}.
To this end, the primary goal is to design quantum structures and encoding 
strategies characterized by ``long'' decoherence times, compared to 
typical time-scale of gating.
The first well-defined semiconductor-based  proposal of QC \cite{DILO}
relies on  spin dynamics in quantum dots (QD), 
it exploits the  low decoherence of spin degrees of
freedom in comparison to the one of charge excitations.
However the proposed manipulation schemes 
are based on  spin dynamics control that
would allow a number of gate operations within the decoherence time
smaller than the one desired by theoretical QC.
On the other hand gating of charge excitations could
be envisioned  by resorting to {\sl present} 
 ultrafast laser technology, that  is now able to generate electron-hole
quantum states on a sub-picosecond time-scale
and to perform
on such states a variety of coherent-carrier-control operations \cite{CCC}.
More specifically, 
this suggests the idea of designing fully optical gating schemes based on 
inter-qubit coupling mechanisms as, e.g., optical 
nonlinearities and dipole-dipole coupling.
In this respect
decoherence times on nano/microsecond scales 
can be regarded as ``long'' ones.\\
\indent Following this spirit, in this letter we investigate a 
semiconductor-based 
implementation of 
the noiseless quantum encoding proposed in \cite{ZR}. 
The idea  is that, in the presence of a sort of  'coherent' environmental noise,
one  can identify states  that are hardly
corrupted rather than states that can be easily corrected.
More specifically, we show that by 
choosing as quantum register an array of quantum dots 
\cite{QD-reviews} 
and by preparing the QD system in proper multi-dot quantum 
states,
it is possible to strongly suppress 
 electron-phonon scattering, 
which is known to be the primary source of decoherence in semiconductors 
\cite{Shah}. 
The physical system under investigation consists of an array of $N$ identical 
quantum dots, whose Hamiltonian can be schematically written as 
${ H} = { H}_c + { H}_p + { H}_{cp}$. 
The term 
${ H}_c = \sum_i { H}_c^i = \sum_{i\alpha}
\epsilon_\alpha a^\dagger_{i\alpha} a^{ }_{i\alpha}
$
describes the non-interacting carrier system, 
$i$ and $\alpha$ being, respectively, the QD index and energy level, while 
$
{ H}_p = \sum_{\lambda{ q}} \hbar\omega_{\lambda{ q}} 
b^\dagger_{\lambda{\bf q}} b^{ }_{\lambda{\bf q}}
$
is the free-phonon Hamiltonian, $\lambda$ and ${\bf q}$ denoting, 
respectively,  the phonon mode and wavevector.
The last term accounts for the coupling of the carriers in the QD 
array with the different phonon modes of the crystal:
%\begin{equation}
$
{ H}_{cp} = \sum_{i\alpha,i'\alpha';\lambda{\bf q}} 
\left[
g^{ }_{i\alpha,i'\alpha';\lambda{\bf q}} 
a^\dagger_{i\alpha} b^{ }_{\lambda{\bf q}} a^{ }_{i'\alpha'} +
\mbox{H.C.} \right].
\label{H_cp}
$
%\end{equation}
Here, 
%\begin{equation}
$
g_{i\alpha,i'\alpha';\lambda{\bf q}} = \tilde{g}_{\lambda{\bf q}}
\int \phi^*_{i\alpha}({\bf r}) 
e^{i{\bf q \cdot r}} \phi^{ }_{i'\alpha'}({\bf r}) d{\bf r}
$
%\label{g}
%\end{equation}
are the matrix elements of the phonon potential between the 
quasi-0D states $i\alpha$ and $i'\alpha'$. The explicit form of the coupling 
constant $\tilde{g}_{\lambda{\bf q}}$ depends on the particular phonon mode 
$\lambda$, e.g., acoustic, optical, etc.
We will assume that only the two lowest energy levels in each
dot ( $\alpha = 0,1$)
will play a role in the quantum-computation dynamics \cite{NOTE1}.
The dynamics of this low-energy sector coupled with the phonon modes of the 
crystal is therefore mapped onto the one of $N$ two-level 
systems ({\it qubits}) linearly coupled with the bosonic 
degrees of freedom $\lambda{\bf q}$, 
the latter representing the decoherence-inducing environment of  
the  computational (i.e., carrier) subsystem.
To describe the so obtained $N$-qubit register it is then convenient to adopt 
the  spin formalism \cite{note-spin}:
$\{ \sigma_i^{\eta}\}_{i=1}^N\,(\eta=\pm,z)$ 
will denote the Pauli matrices 
spanning $N$ {\it local} $sl(2)$ algebras.
It is also convenient to introduce the {\it global} 
$sl(2)$ algebra generated by the collective spin operator
$S^\eta = \sum_i \sigma_i^\eta,\,(\eta=\pm,z)$.
We will assume  no direct phonon coupling 
between different qubits, i.e.,
$g_{i\alpha,i'\alpha';\lambda{\bf q}} = 0$ for $i \ne i'$;
In order to obtain a closed-form equation for the reduced density matrix $\rho$
describing our qubit array,
one can trace out the phonon degrees of freedom by means of the standard 
Born-Markov approximation \cite{PZ}. 
The resulting ``master equation'' is of the form 
$\dot\rho={\cal L}(\rho)$.
Here, the Liouvillian superoperator ${\cal L}$ is given by the sum of 
two contributions:
a unitary part ${\cal  L}_u,$ that preserves quantum coherence, 
plus a dissipative one 
${\cal L}_d,$ describing
irreversible decoherence-dissipation processes.
To better understand such separation, 
it is useful  to introduce the hermitian matrix $\Delta_{ii'}^{(\pm)}$ 
($\Gamma_{ii'}^{(\pm)})$
as the real (immaginary) part of the matrix
\begin{equation}
L_{ii'}^{(\pm)}=\sum_{\lambda{\bf q}}\frac{g_{i,\lambda{\bf q}}\bar 
g_{i',\lambda{\bf q}}} {
E - \hbar\omega_{\lambda{\bf q}} - i \,0^+} 
\left(n_{\lambda{\bf q}}+ \theta(\mp)\right)
\label{Matrix} ,
\end{equation}
where $g_{i,\lambda{\bf q}}\equiv g_{i 1,i 0, \lambda{\bf q}}$ 
and $E=\epsilon_1-\epsilon_0.$ 
Here, $n_{\lambda{\bf q}}$ and $\theta$ are respectively the 
Bose thermal distribution and the step function.
As we will see, this matrix encodes the spatial correlations
of the quantum register defining the {\sl effective topology} that 
can be probed
by the phononic environment.
In particular, the spectral  data of  ${\bf L}^{(\pm)}$ contain
 information about the existence of {\sl subdecoherent} subspaces \cite{PZ}.    
More specifically, one finds  
${\cal L}_u (\rho)={i/\hbar}\,[\rho, { H}_c+\delta{ H}_c]$ 
where the phonon-induced 
renormalization $\delta{ H}_c$ to our free-qubit Hamiltonian ${ H}_c$
[which in our spin formalism reads
${ H}_c=\epsilon \,S^z$] is given by
$
\delta{ H}_c =\sum_{\eta=\pm}
\sum_{ ii'=1}^N \Delta_{ii'}^{(\eta)} \sigma^{-\eta}_i\,\sigma^{\eta}_{i'}
$
These contributions are usually referred to as the Lamb-shift term 
In contrast, 
the dissipative (non-unitary) component of the Liouvillian is given by 
${\cal L}_d(\rho)=\sum_{\eta=\pm}{\cal L}^{\eta}_d(\rho)$ where
the emission ($\eta=-$) and the absorption ($\eta=+$) terms can be cast
into the compact form
\begin{equation}
{\cal L}^{\eta}_d(\rho)=\frac{1}{\hbar}\sum_{ii',\eta=\pm} 
\Gamma^{(\eta)}_{ii'}\, \left (
[\sigma_i^\eta 
\,\rho,\,\sigma_{i'}^{-\eta}]+
[\sigma_i^\eta,\,\rho\,\sigma_{i'}^{-\eta}]\right ).
\end{equation}
The diagonal ($i = i'$) terms describe the usual carrier-phonon 
scattering processes in a single quantum dot, 
 as obtained 
from Fermi's golden rule; 
In contrast, the off-diagonal  elements are not 
positive-definite and describe collective coupling effects, which play a 
crucial role in the realization of a decoherence-free evolution \cite{ZR}.

In order to study the corruption of the information encoded in the initial
{\sl pure} preparation $\rho=|\psi\rangle\langle\psi|$ 
it is useful to introduce
the following quantity: the {\sl fidelity} 
$F(t)\equiv \langle\psi|\rho(t)|\psi\rangle.$
We define the first order decoherence time (rate) 
$\tau_1$ ($\tau_1^{-1}$) 
in terms of the short-time expansion 
$F(t)=1-t/(\tau_1) +o(t^2).$ 
If $|\psi\rangle$ is a total spin eigenstate, it is easy to check that
$\hbar\,\tau_1^{-1}[|\psi\rangle]= 
\langle\psi | { H}_{eff}|\psi\rangle$
where the effective Hamiltonian ${ H}_{eff}$ has the  same structure of 
the Lamb-shift term $\delta H_c$
with ${\bf \Gamma}$ replacing ${\bf \Delta}$ [see Eq.~(\ref{Matrix})].
Notice that: 
i) $\tau_1\ge 0$ (i.e., $H_{eff}$ is a positive operator);  
ii) in this (first-order) decoherence time the Lamb-shift
terms {\sl do not} play any role.

To exemplify the collective nature of the decoherence process let us consider
the decoherence rate
for the states 
 $|\psi_{\cal D}\rangle\equiv 
\otimes_{(i,i')\in{\cal D}}( |01\rangle-|10\rangle)_{ii'}$
(here, ${\cal D}$ is a dimer partition of the qubit array)
that are {\sl singlets} of the global $sl(2)$ algebra \cite{NOTE2}.
In this case one gets  
$\tau_1^{-1}= \sum_{\eta=\pm} 
(2\,\tau_\eta)^{-1} \,f_{\cal D}({\bf{\Gamma}}^{(\eta)})$,
where 
$\hbar\,\tau^{-1}_\eta={N}\,\Gamma^{(\eta)}_{11}$
 is the (maximal) decoherence rate for $N$ uncorrelated qubits and
\begin{equation}
f_{\cal D}({\bf \Gamma})= 1-\frac{2}{N}\, \Re\sum_{(i,i')\in{\cal D}} \Gamma_{ii'}/\Gamma_{11}.
\end{equation}
The quantity $f_{\cal D}$ contains the information about the degree of 
multi-qubit correlation
in the decay process.
Suppose now that one is able to design our qubit array in such a 
way that
$f_{\cal D}({\bf \Gamma})= 0$ then $1/\tau_1=0$ that means that our coding state
$|\psi\rangle$ is on a short time scale unaffected by decoherence;
moreover if $|\psi\rangle$ is annihilated by ${\cal L}_u$ as well 
it turns out to be {\sl noiseless}: it does not suffer any evolution at all.
Generally speaking there are
two extreme cases in which ${ H}_{eff}$ is easily diagonalized.
i) If ${\bf{\Gamma}}^{(\eta)}\propto {\bf{I}}$  then $H_{eff}$ has a 
trivial kernel and the
 qubits decohere independently: no subdecoherent encoding exists.
In this limit  the environment ``sees'' a register endowed  with a 
discrete topology.
The same is true for an initial {\sl unentangled } preparation (i.e., simple tensor
product).
ii) $\Gamma_{ii'}^{(\pm)}= const$ 
the register gets ``point-like''.
In this case, the effective Hamiltonian is  bilinear in the $S^\eta$'s 
and the subdecoherent subspace coincides with singlet sector of the global 
$sl(2)$ algebra \cite{ZR}.

%%%%
The above theoretical analysis has been applied to state-of-the-art 
quasi-0D semiconductor heterostructures.
In particular, a linear array of vertically stacked quantum dots ---along 
the growth ($z$) direction--- has been considered;
More specifically, the array is formed by GaAs/AlGaAs structures 
similar to that of \cite{QD-str} alligned on the same $z$ axis.
%%%%
The three-dimensional confinement potential 
giving rise to the quasi-0D single-particle states 
$\phi_\alpha$ is properly described in terms of a quantum-well (QW) profile 
along the growth direction times a two-dimensional (2D) 
parabolic potential in the normal plane.
Since the width $d$ of the GaAs QW region is typically of the order of few 
nanometers, the energy splitting due to the quantization along the growth 
direction is much 
larger than the confinement energy $E$ induced by the 2D parabolic potential
(typically of a few meV).
Thus, the two single particle states ---state $|0\rangle$ and
$|1\rangle$--- realizing the {\sl qubit} considered  
 so far are given by products of 
the QW ground state times the ground or first excited state of the 2D 
parabolic potential \cite{note-degeneracy}, 
their energy splitting  being 
equal to $E$.
\begin{figure}
\unitlength1mm
\ \par\noindent
\begin{picture}(80,55)
\put(0,0){\psfig{figure=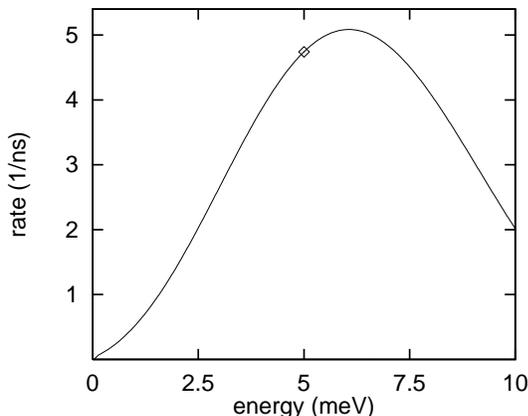,width=80mm}}
\end{picture}
\caption{\label{fig1}
%%%%
Carrier-phonon scattering rate for a single QD structure as a function of
the energy splitting $E$ at low temperature (see text).
}
\end{figure}

\begin{figure}
\unitlength1mm
\ \par\noindent
\begin{picture}(80,55)
\put(0,0){\psfig{figure=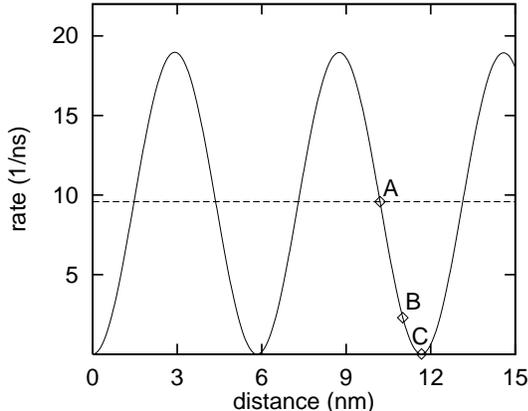,width=80mm}}
\end{picture}
\caption{\label{fig2}
Phonon-induced decoherence rate
%[see Eq.~\protect\ref{f_sing})]
for a four-QD array (solid line) as a
function of the inter-dot distance $a$ compared with the corresponding
single-dot rate (see text).
}
\end{figure}

\begin{figure}
\unitlength1mm
\ \par\noindent
\begin{picture}(80,55)
\put(0,0){\psfig{figure=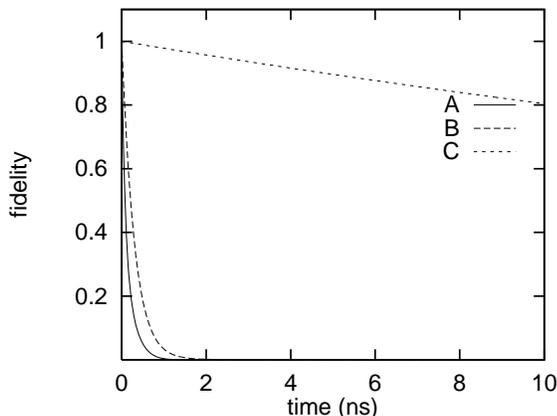,width=80mm}}
\end{picture}
\caption{\label{fig3}
Fidelity $F$ as a function of time as obtained from
a direct numerical solution of the Master equation for the relevant case of
a four-QD array (see text).
}
\end{figure}

As a starting point, let us discuss the role of carrier-phonon interaction 
in a single QD structure with $d=4$\,nm.
Figure \ref{fig1} shows the total (emission plus absorption) carrier-phonon 
scattering rate 
($\Gamma^+_{ii}+ \Gamma^-_{ii}$) 
at low temperature ($T = 10$\,K)
as a function of the energy spacing $E$. 
Since the energy range considered is smaller than the optical-phonon energy
 ($36$\,meV in GaAs ), due to energy conservation scattering with LO phonons
is not allowed. Therefore, the only phonon mode
$\lambda$ which 
contributes to the rate of Fig.~\ref{fig1} is that of acoustic 
phonons. 
The latter has been evaluated starting from the explicit form of the 
carrier-phonon matrix elements $G$ which, in turn, involve the 3D wave 
functions as well 
as the explicit form of the deformation-potential coupling $\tilde{g}$. 
%%%%
To this end, a bulk phonon model in the long-wavelength limit has been 
employed \cite{note-phonons}.
%%%%
Again, due to energy conservation, the only phonon wavevectors 
involved must satisfy
$|{\bf q}| = {E/\hbar c_s}\equiv q$, 
$c_s$ being the GaAs sound velocity. It follows that by increasing the 
energy spacing  $E$ the wavevector $q$ is increased, 
which reduces the carrier-phonon 
coupling $g$ entering in the electron-phonon
interaction and then  the scattering rate. 
%%%%
This well-established behaviour is typical of a quasi-0D structure. 
As shown in Fig.~\ref{fig1}, for 
$E = 5$\,meV ---a standard value for many state-of-the-art QD structures---
the carrier-phonon scattering rate is 
significantly suppressed compared to typical bulk values \cite{Shah}.\\
\indent We will now show that by means of a proper information encoding, i.e., a 
proper choice of the initial multi-system quantum state, and a proper design
of our QD array,
we can strongly  suppress  phonon-induced decoherence processes, thus further 
improving the 
single-dot scenario discussed so far.
Let us now consider a four-QD array 
(i.e., the simplest noiseless qubit register).
>From the short-time 
expansion discussed above, we have numerically evaluated the decoherence 
rate for such QD array choosing as energy splitting $E = 5$\,meV (see 
Fig.~\ref{fig1}). As  initial state we have chosen 
the singlet defined by the dimer
partition ${\cal D}_1=\{(1,2),\,(3,4)\}$.
The resulting decoherence rate is shown as solid line in Fig.~\ref{fig2} as 
a function of the inter-dot distance $a$. 
The uncorrelated-dot decoherence rate is
also reported as dashed line for comparison.
Rather surprisingly, in spite of the 3D nature of the sum over ${\bf q}$ 
entering the calculation of the function $\Gamma^{(\pm)}_{ii'}$ 
[see Eq.~(\ref{Matrix})], 
the decoherence rate exhibits a periodic behaviour
over a range comparable to the typical
QD length scale.
This effect  -- which 
would be the natural for a 1D phonon 
system -- 
 stems from the exponential suppression, in the overlap integral,
of the contributions of phononic modes with non-vanishing in-plane 
component. 
The 1D  behavior 
is extremely important since it allows, by suitable choice of
the inter-dot distance $a$, to realize the symmetric regime ii)
in which all the dots experience the {\sl same} phonon field and therefore 
decohere collectively. 
Indeed, by taking $\bar a= n\,2\pi/q \,(n\in {\bf {N}})$
one finds, for example, that $f_{{\cal D}_1}(\bar a) \ll 1$ 
Figure \ref{fig2} shows  that for the particular QD structure considered,
case C should correspond to a decoherence-free evolution of
a singlet state, which is not the case for A and B (see simbols
in the figure).
In order to extend the above short-time analysis, we have performed a full
time-dependent solution 
by direct integration 
of the Master equation for the density matrix $\rho$,
taking also in to account the Lamb-shift terms. 
Starting from the same GaAs QD structure considered so far, we have 
simulated the above noiseless encoding for a four-QD array. 
Figure \ref{fig3} shows the fidelity as a 
function of time as obtained from our numerical solution of the Master
equation. In particular, we have performed three different simulations 
corresponding to the different values of $a$ depicted in Fig.~\ref{fig2}.
Consistently  with our short-time analysis, 
for case C we find a strong suppression of the 
decoherence rate which extends the sub-nanosecond time-scale of the  B 
case (corresponding to twice the single-dot rate) to the microsecond time-scale.
This confirms that by means of the proposed encoding strategy one can realize a 
decoherence-free evolution over a time-scale comparable with typical 
recombination times in semiconductor materials \cite{Shah}.\\
\indent 
At this point a few comments are in order. 
The actual implementation of the suggested encoding relies, of course,
on precise quantum state sinthesis and manipulations.
This crucial point  -- that was not the focus of this paper --
might be addressed, by resorting for example to the ideas 
of the early QD proposal in \cite{BEDJ}.  
Moreover it is well-known that carrier-phonon scattering is not 
the only source of 
decoherence in semiconductors. In conventional bulk materials 
also carrier-carrier interaction is found to play a crucial role. However, 
state-of-the-art QD structures ---often referred to as semiconductor 
macroatoms \cite{QD-reviews}--- 
can be regarded as few-electron systems basically decoupled from the 
electronic degrees of freedom of the environment. 
For the semiconductor QD array considered,
the main source of Coulomb-induced ``noise'' may arise from the inter-dot 
coupling. However, since such Coulomb coupling vanishes for large values of 
the QD separation and since the proposed encoding scheme 
can be realized for values of $a$ much larger than the typical 
Coulomb-correlation length (see Fig.~\ref{fig2}), 
a proper design of 
our quantum register may  rule out 
 such additional decoherence channels.

In summary, we have investigated a semiconductor-based implementation of
a quite general quantum-encoding strategy, which allows to suppress 
phonon-induced decoherence on the carrier subsystem. 
More specifically, we have shown that an array of state-of-the-art QD 
structures is a suitable  qubit register since it allows to realize 
a decoherence-free evolution on a  time-scale long compared with 
those of modern ultrafast laser-pulse generation and manipulation. 
Since the latter is the natural candidate
for quantum gating of charge excitations in semiconductor nanostructures, 
this result might constitute an important first step toward
a solid-state implementation of quantum computers. 

We are grateful to M. Rasetti  for stimulating and fruitful discussions. 
This work was supported in part by the EC Commission through the TMR Network 
``ULTRAFAST''. P.Z. thanks Elsag-Bailey for financial support.

\end{multicols}

\begin{references}
\bibitem{NOTE1} 
For $E\gg k_B\,T$ the {\sl leakage} errors induced
by unwanted transitions to higher excited states is negligible.
\bibitem{note-spin}
Despite the spin formalism we are concerned with
 the charge degrees of freedom of the carrier subsystem only.
\bibitem{note-degeneracy}
With ``first excited state'' here we mean any linear combination of the two 
energetically degenerate states.
\bibitem{note-phonons}
Within this bulk model various phonon-confinement effects, 
e.g., interface phonon modes, are neglected. However, while these effects 
may affect significantlythe phononic energy spectrum, they are by far less 
decisive in the determination of the carrier-phonon coupling $g$ 
\protect\cite{Shah}\protect.
\end{references}
\end{document}